\documentclass{webofc}
\usepackage[varg]{txfonts}   
\usepackage{subcaption}
\usepackage{xcolor}
\begin{document}
\title{Overview of ALICE results in pp, pA and AA collisions}
%
%

\author{\firstname{Rainer} \lastname{Schicker, for the ALICE Collaboration} \inst{1}\fnsep\thanks{\email{schicker@physi.uni-heidelberg.de}}
}

\institute{Phys. Inst., Im Neuenheimer Feld 226, 69120 Heidelberg}

\abstract{

The ALICE experiment at the Large Hadron Collider (LHC) at CERN is 
optimized for recording events in the very large particle multiplicity 
environment of heavy-ion collisions at LHC energies. The ALICE collaboration 
has taken data in Pb-Pb collisions in Run I and Run II  at nucleon-nucleon 
center-of-mass energies $\sqrt{s_{\text{NN}}}$ = 2.76 and \mbox{5.02 TeV},
respectively,  and in pp collisions at center-of-mass energies 
$\sqrt{s}$ = 0.9, 2.76, 5.02, 7, 8 and 13 TeV. The asymmetric system p-Pb 
was measured at a center-of-mass energy 
\mbox{$\sqrt{s_{\text{NN}}}$ =  5.02 TeV.} 
Selected physics results from the analysis of these data are presented, 
and an outline of the ALICE prospects for Run III is given. 
}
\maketitle
\section{Introduction}
\label{intro}

The ALICE experiment at the Large Hadron Collider (LHC) at CERN is a  
general purpose detector optimized for the measurement of high-energy 
Pb-Pb collisions \cite{Alice1}. The analysis of such collisions allows to 
study a variety of issues which are pertinent to an improved understanding 
of Quantum Chromodynamics (QCD), the theory of the strong interaction. 
At high temperatures and energy densities of the QCD medium, lattice QCD 
calculations predict a phase transition to a state of deconfined quarks 
and gluons \cite{LQCD}. High-energy nuclear collisions allow to reach such 
energy densities, however within a finite volume and for a limited time only. 
This phase transition is accompanied by the restoration of chiral symmetry, 
in which the quarks acquire their current mass. The understanding of the 
properties of the QCD medium created in heavy-ion collisions  
necessitates a good knowledge of the underlying collision dynamics. 
Intrinsic QCD-medium signals can be disentangled from initial
cold-matter and final-state effects by comparing the heavy-ion
observables to the corresponding quantities measured in 
pp and p-Pb collisions.

\section{The ALICE Experiment}
\label{sec-1}

\subsection{The History of the ALICE Experiment}

The first conceptual ideas for experiments at the LHC were discussed in a 
workshop in 1990. Out of this workshop, an Expression of Interest, followed
by a Letter of Intent and the Technical Proposal evolved.
After approval of the ALICE central barrel system  in 1997, the  
different detector systems were specified in Technical Design Reports.  
A muon spectrometer was later added, as well as a Transition
Radiation Detector (TRD) and Electromagnetic Calorimeters EMCAL and DCAL 
in the central barrel.

\subsection{The ALICE Physics Programme}

\subsubsection{The ALICE Physics Programme in pp-Collisions}
\vspace{-0.1cm}
The ALICE physics programme in pp collisions is complementary to the other LHC
experiments due to the low p$_{T}$-threshold, and due to the excellent particle
identification capability \mbox{in the central barrel \cite{AlicePerf}.} 
Besides being a reference for the analysis of Pb-Pb collisions, the 
pp collisions have interest in their own. The analysis of 
minimum-bias pp collisions addresses a plethora of issues relevant
for the understanding of soft QCD processes, such as particle and 
multi-particle production in the non-perturbative QCD regime.
The analysis of high-multiplicity pp collisions allows to study the 
possible onset  of collectivity in small systems, such as evidenced
in collective flow effects.
      
\vspace{-0.2cm}
\subsubsection{The ALICE Physics Programme in pA-Collisions}
\vspace{-0.1cm}
The analysis of data taken in p-Pb collisions, and its comparison
to the corresponding quantity measured in Pb-Pb collisions, can be 
used to disentangle cold-matter initial and final-state effects from 
intrinsic QCD-medium effects. Of particular interest here 
are possible shadowing of nuclear Parton Distribution Functions (nPDFs),
and the low-x behaviour of the gluon-PDF as signature of 
gluon saturation at low values of Bjorken-$x$. 

\vspace{-0.2cm}
\subsubsection{The ALICE Physics Programme in AA-Collisions}
\vspace{-0.1cm}
The analysis of Pb-Pb collisions at LHC energies addresses a multitude of 
physics issues, above all the question of experimental observables to 
characterize the nature of the QCD phase diagram \cite{PBMWam}. 
The observables derived in these analyses, and their correlations, can be 
used to study experimental signatures of the deconfinement and the chiral 
symmetry transition, and to examine the relationship between these 
two transitions.

\subsection{The ALICE Detector Systems}
\label{sec-2}

\begin{figure}[ht]
\centering
\sidecaption
\includegraphics[width=10.2cm]{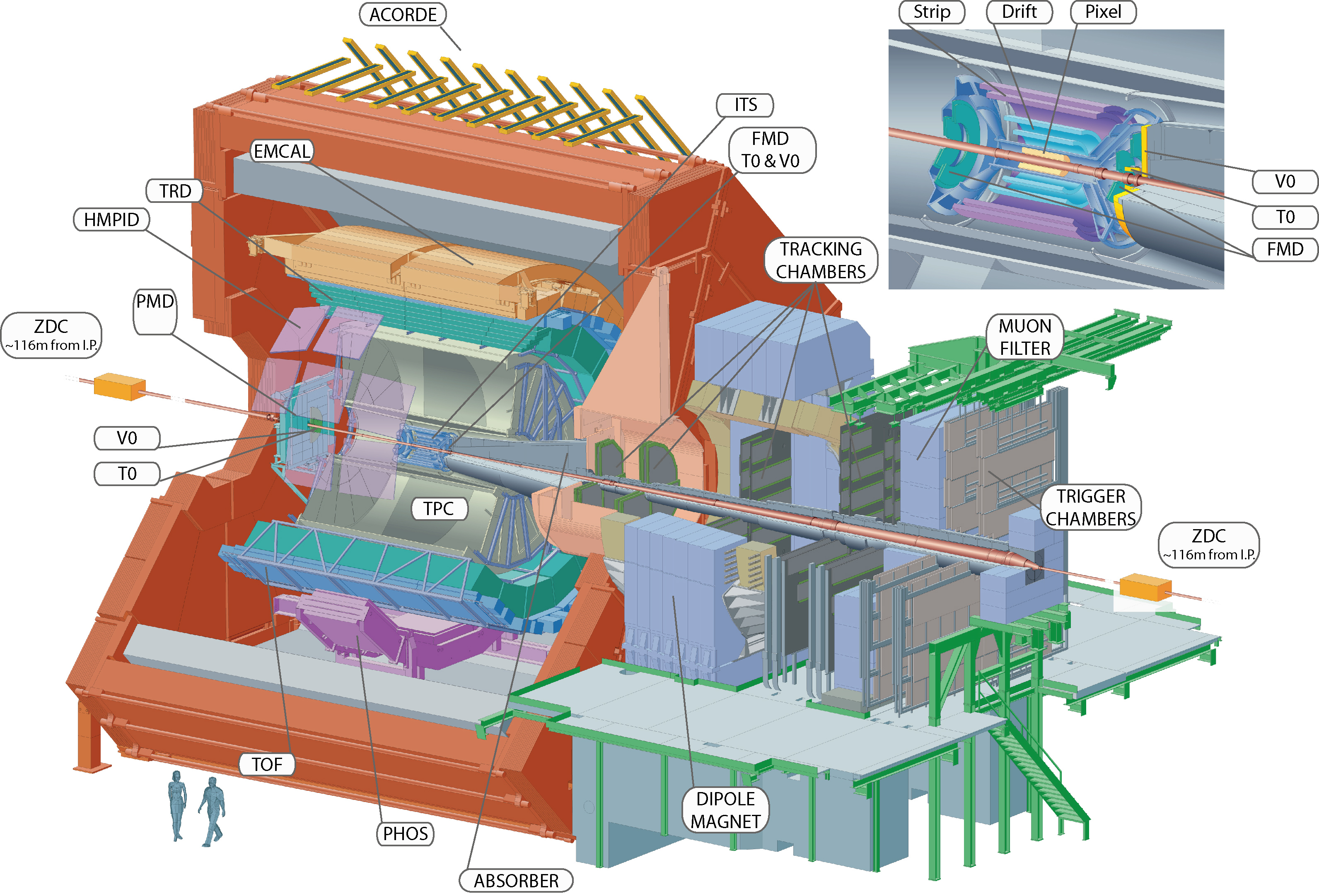}
\caption{The detector systems of the ALICE central barrel, with the muon 
spectrometer shown on the right hand side of the central barrel.}
\label{fig-1}       
\end{figure}

\vspace{-0.1cm}
In Fig. \ref{fig-1}, the yoke of the ALICE magnet is shown in red. The 
innermost detector system, the Inner Tracking System (ITS), is zoomed in on 
the top right of the figure, showing the strip, drift and pixel subsystems. 
Radially outwards from the ITS, the Time Projection Chamber (TPC), the 
Transition Radiation Detector (TRD), and the Time-of-Flight detector (TOF) 
are shown. The V0 system serves as trigger detector, and its information
is used to define event parameters such as high-multiplicity or
rapidity-gap characteristics.  In addition, there exist other small
acceptance systems for particle identification and event characterisation
(FMD, HMPID, T0). The electromagnetic calorimeters EMCAL and PHOS are able 
to measure photons in a limited solid angle of the central barrel.
A muon spectrometer complements the central barrel on one
side within the pseudorapidity range $-4 < \eta < -2.5$.

\subsection{The ALICE Performance}
\label{sec-3}

The ALICE detector systems shown in Fig. \ref{fig-1} provide the information
for reconstructing the individual tracks, for fitting their momentum, and for
identifying the particle type of the track. 

\begin{figure}[ht]
\begin{center}
\includegraphics[width=6.8cm]{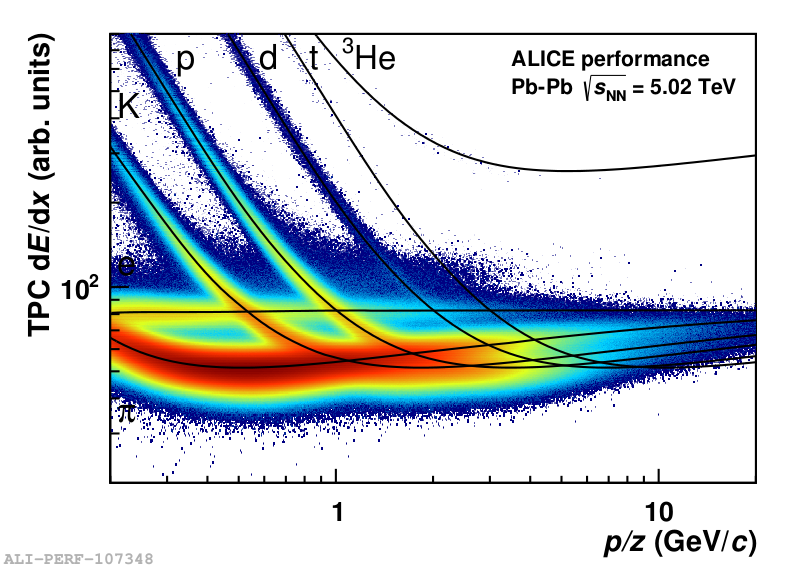}
\includegraphics[width=7.2cm]{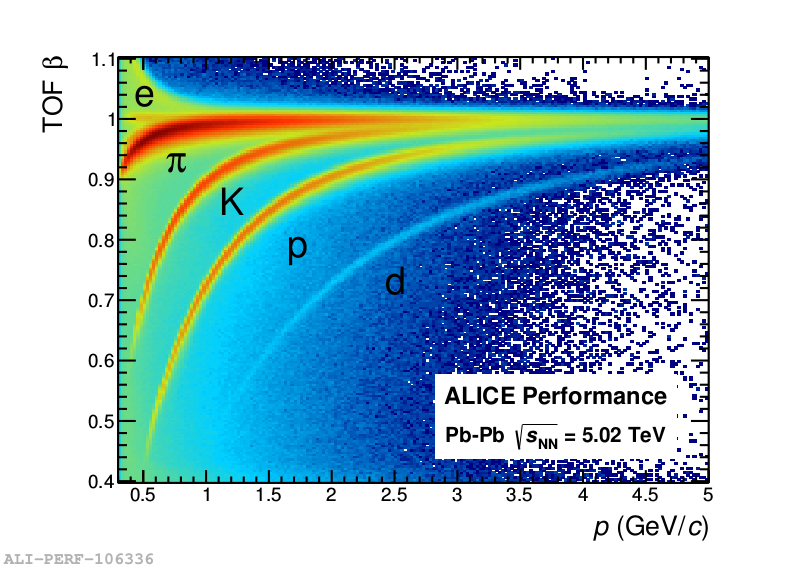}
\caption{The particle identification by energy loss dE/dx in the TPC is shown 
on the left. The particle identification by TOF is shown on the right.}
\label{fig-2}       
\end{center}
\end{figure}

The ALICE particle identification capability by the energy loss dE/dx in the
TPC is shown on the left part, and by the Time-of-Flight as measured by the 
TOF detector on the right part of  Fig. \ref{fig-2}. The information from the 
TRD detector, located between the TPC  and the TOF detector, is used to 
identify  electrons. The separation of electrons from charged hadrons 
can in addition be achieved by the electromagnetic calorimeters.

\subsection{The Data Sample collected by ALICE}
\label{sec-4}

The statistics of the data samples recorded by the ALICE collaboration 
in Run I and Run II of the LHC is displayed in Table \ref{tab-1}.

\begin{table}[ht]
\centering
\caption{LHC Runs I+II, 2009-2015:}
\label{tab-1}       
\begin{tabular}{lllll}
\hline
system & run & year & $\sqrt{s_{\text{NN}}}$ (TeV) & int. lumi.  \\\hline
p-p & I & 2009-2010 & 0.9 & 0.15 nb$^{-1}$ \\
p-p & I & 2011 & 2.76 & 1.1 nb$^{-1}$ \\
p-p & I & 2010-2011 & 7 & 4.8 pb$^{-1}$ \\
p-p & I & 2012 & 8  & 9.7 pb$^{-1}$ \\
p-p & II & 2015 & 5.02 & 2.5 nb$^{-1}$ \\
p-p & II & 2015 & 13 & 4.35 pb$^{-1}$ \\
p-Pb & I & 2013 & 5.02 & 30 nb$^{-1}$ \\
Pb-Pb & I & 2010,2011 & 2.76 & 0.1 nb$^{-1}$ \\
Pb-Pb & II & 2015 & 5.02 & 0.24 nb$^{-1}$ \\\hline
\end{tabular}
\end{table}

In Run I and Run II, the ALICE collaboration has taken data in
pp collisions at center-of-mass energies of 
$\sqrt{s}$ = 0.9, 2.76, 5.02, 7, 8 and 13 TeV. In Pb-Pb collisions, 
events were recorded in Run I and Run II at center-of-mass energies 
of $\sqrt{s_{\text{NN}}}$ = 2.76 and 5.02 TeV, respectively. In p-Pb
collisions, data were recorded in Run I at the center-of-mass energy
of $\sqrt{s_{\text{NN}}}$ = 5.02 TeV. During November and December of 
this year, data taking in p-Pb collisions is planned
at $\sqrt{s_{\text{NN}}}$ = 5.02 and 8 TeV.

\section{Proton-Proton Collisions}
\label{sec-5}

\subsection{Multi-Parton Interactions}

Particle production in high-energy hadronic collisions can get substantial
contributions from Multi-Parton Interactions (MPI). Here, several partonic
interactions can take place in a hadronic collision, resulting in a 
multiplicity dependence of observables such as particle yields and 
transverse momentum distributions. Analysing MPI effects is of fundamental 
importance for studying the range of validity of Parton Distribution 
Functions (PDFs), an input when calculating perturbative QCD cross sections.

\begin{figure}[h!]
\begin{subfigure}[t]{0.48\textwidth}
\begin{center}
\includegraphics[width=4.8cm]{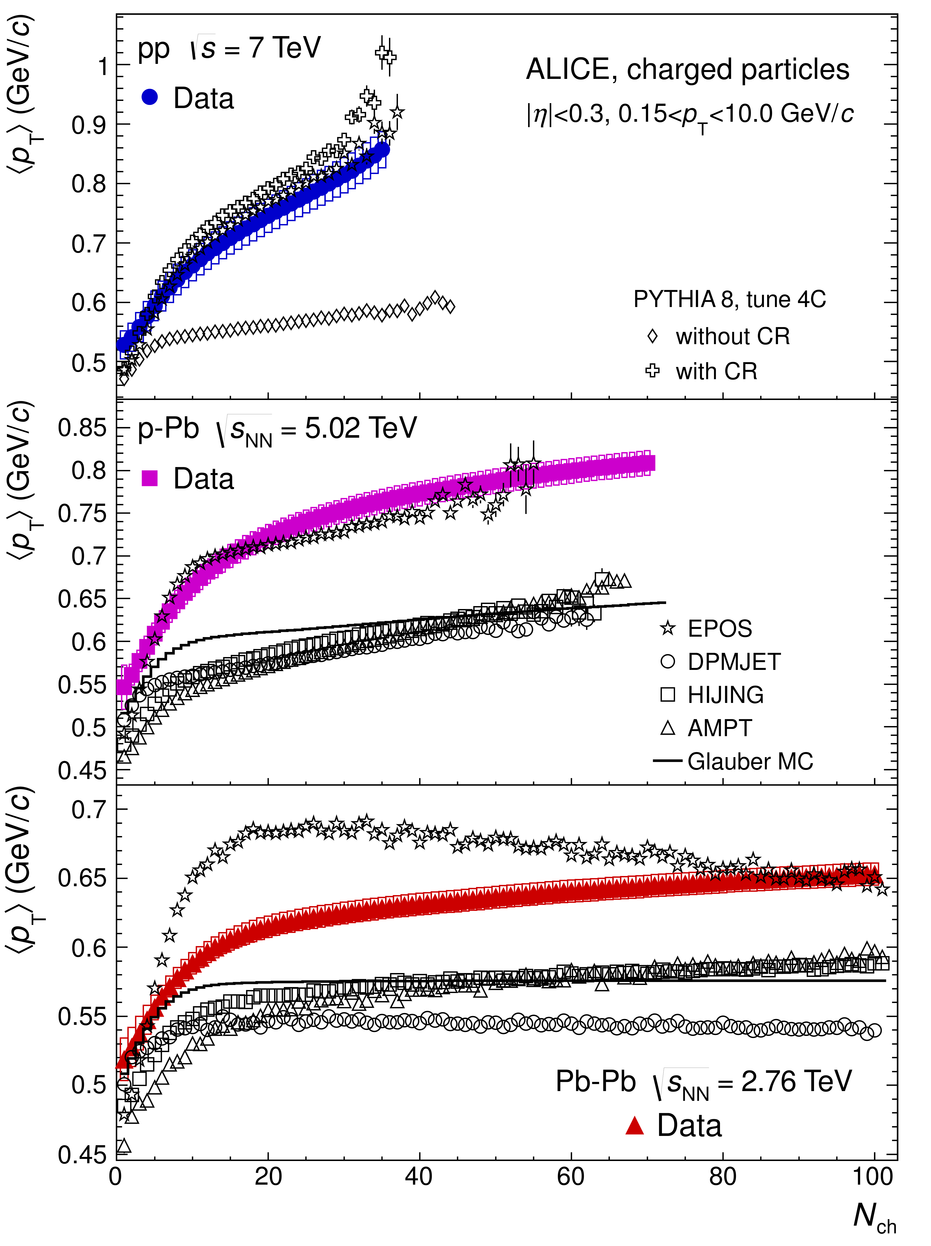}
\end{center}
\end{subfigure}
\begin{subfigure}[ht]{0.50\textwidth}
\vspace{-5.0cm}
\begin{center}
\includegraphics[width=4.8cm]{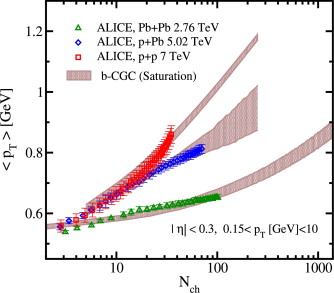}
\end{center}
\end{subfigure}
\caption{Average transverse momentum $<\!\!p_{T}\!\!>$ as function of 
charged particle multiplicity for pp, p-Pb and Pb-Pb collisions are shown on 
the left. The comparison of measured $<\!\!p_{T}\!\!>$ to the predictions 
of the Color Glass Condensate  approach are shown on the right.}
\label{fig-3}       
\end{figure}

\vspace{-0.4cm}
In the top left panel in Fig. \ref{fig-3}, the ALICE measurement of the 
average transverse  momentum in the range $|\eta| <$ 0.3 for pp collisions 
at $\sqrt{s}$ = 7 TeV is displayed as function of the charged particle 
multiplicity \cite{AliceMPI}. The results of the PYTHIA 8 calculations, 
shown by the open symbols, clearly show the importance of including colour 
reconnections. In the middle panel on the left, the corresponding ALICE 
data  for p-Pb collisions are shown in purple, superimposed in black by the 
predictions  of the EPOS, DPMJET, HIJING and AMPT models \cite{EvGen}. 
The EPOS model predicts the ALICE data reasonably well, even though it 
shows a different trend at multiplicities N$_{ch} <$ 20 \cite{EPOS1}. 
None of the other three models is able to describe the data. A Glauber 
MC calculation, shown by the solid black line, also fails to reproduce the 
data. In this Glauber MC approach, p-Pb collisions are assumed to consist of 
independent pp collisions, each contributing to the particle yield according 
to the yield as measured in pp collisions. In the bottom panel on the left, 
the ALICE data  for Pb-Pb  collisions are shown in red, superimposed in black 
by the predictions  of the different model calculations. 
As for the p-Pb data, the DPMJET, HIJING and AMPT model  as well as the MC 
Glauber calculation underpredict the measured $<\!\!p_{T}\!\!>$.
The EPOS model overpredicts the data and shows a different trend.
 
On the right side of Fig. \ref{fig-3}, the ALICE data are compared to 
Color Glass Condensate (CGC) model calculations \cite{CGC}. The CGC 
approach, addressing the importance of the initial-state effect,  is based on 
gluon saturation at low values of Bjorken-$x$, and Glasma physics. The CGC 
predictions are in much better agreement with the ALICE data than the 
model predictions shown on the left of Fig.  \ref{fig-3}.

\subsection{Strangeness and Charm in Proton-Proton Collisions}

The initial state of proton-proton collisions does not contain strange
valence quarks. The measurement of strange hadrons hence allows the study 
of particle production in QCD both in the perturbative and non-perturbative
sector. Strangeness can be created in hard perturbative $2 \rightarrow 2$ 
partonic scatterings by flavour creation 
$(gg \rightarrow s\bar{s}, q\bar{q} \rightarrow s\bar{s})$ and
by flavour excitation ($gs \rightarrow  gs, qs \rightarrow qs$). 
During partonic evolution, gluon splitting $( g \rightarrow s\bar{s})$
contributes to strangeness production. 
At low transverse momenta, strangeness production is dominated
by non-perturbative processes such as \mbox{string fragmentation.}

\begin{figure}[h!]
\begin{subfigure}[t]{0.48\textwidth}
\begin{center}
\includegraphics[width=5.cm]{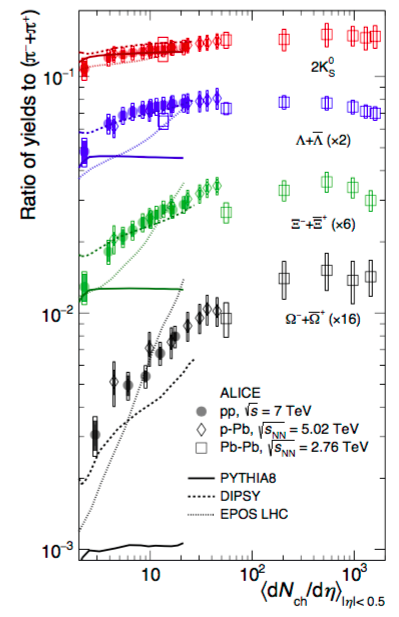}
\end{center}
\end{subfigure}
\begin{subfigure}[ht]{0.50\textwidth}
\vspace{-7.2cm}
\begin{center}
\includegraphics[width=4.8cm]{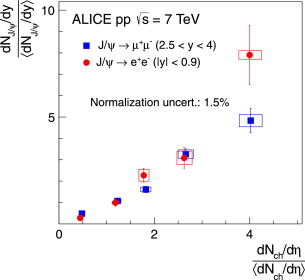}
\includegraphics[height=.02cm,width=6.2cm]{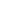}
\includegraphics[width=4.8cm]{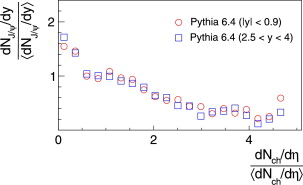}
\end{center}
\end{subfigure}
\caption{The yield ratio of strange and multi-strange hadrons 
to pions as function of (dN/d$\eta)$ is shown on the left.
The normalized J/$\Psi$ yield as function of normalized dN$_{ch}$/d$\eta$ is
shown in the upper part on the right. The PYTHIA prediction of this 
normalized yield is shown in the lower \mbox{part on the right.}}
\label{fig-4}       
\end{figure}

On the left of Fig. \ref{fig-4}, the p$_{T}$-integrated yield 
ratios of strange ($K_{S}^{0},\Lambda,\overline{\Lambda}$) and multi-strange 
($\Xi^{-},\overline{\Xi^{+}},\Omega,\overline{\Omega^{+}}$) hadrons measured 
in pp, p-Pb and Pb-Pb are  shown \cite{Alicestrange}. These ratios represent 
a production rate which increases faster for strange hadrons than for 
non-strange hadrons.  The strangeness enhancement increases with hadron 
strangeness, and with event activity defined by dN$_{ch}$/d$\eta$. This
enhancement is not reflected by the PYTHIA generator
shown in Fig. \ref{fig-4}  by the solid lines.
The EPOS generator describes the trend vs multiplicity,
even though it fails for the triply strange $\Omega$-Baryon.
The Dipsy model describes the data best, but is also not doing well
for the triply $\Omega$-Baryon. 

On the top right part of Fig. \ref{fig-4}, the yield dN$_{J/\Psi}$/dy,
normalized to $<$dN$_{J/\Psi}$/dy$>$, is shown as function of the normalized 
charged particle multiplicity (dN$_{ch}$/d$\eta$)/$<$dN$_{ch}$/d$\eta >$.
This yield is shown in red at midrapidity  |y| $<$ 0.9,
and in blue at forward rapidities 2.5 $<$ |y| $<$ 4.0. 
An approximate linear correlation of these quantities is seen 
in Fig. \ref{fig-4}.  In the bottom right part, the normalized 
dN$_{J/\Psi}$/dy yield as predicted by PYTHIA 6.4 in Perugia 2011 tune is shown.
In this calculation, only J/$\Psi$'s produced in hard scatterings via
the NRQCD framework are calculated, whereas J/$\Psi$ production from 
cluster formation during parton shower evolution or in MPI are not taken 
into account.

\subsection{Charmed Mesons in Proton-Proton Collisions}

The study of open-charm meson production in pp collisions at LHC energies 
allows to test predictions of perturbative QCD (pQCD) at the highest collider 
energies available. Such pQCD calculations are done in collinear factorisation 
approach at next-to-leading order in the general-mass variable-flavour-number 
scheme (GM-VFNS), or at fixed order with next-to-leading-log 
resummation (FONLL). These calculations describe the p$_{T}$-differential
production cross sections of D-mesons within uncertainties.

\vspace{-.2cm}
\begin{figure}[h!]
\begin{center}
\includegraphics[width=7.2cm]{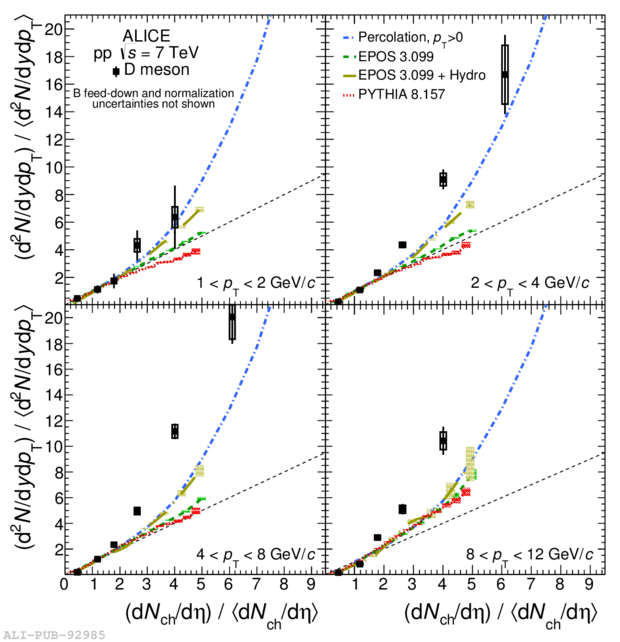}
\end{center}
\caption{Average D-meson relative yield as function of relative 
charged-particle multiplicity at central rapidity in different 
p$_{T}$-intervals.}
\label{fig-5}       
\end{figure}

\vspace{-.4cm}
The relative D-meson yield (average of D$^{0}$, D$^{+}$, D$^{*+}$) is shown in 
Fig. \ref{fig-5} in four p$_{T}$-intervals as function of the relative 
charged particle yield \cite{Alicecharm}. These measurements are compared to 
model predictions by PYTHIA 8 (red-dotted line), the EPOS models (green-dashed 
and green dot-dashed line), and the percolation model (blue dot-dashed line). 
In PYTHIA 8, heavy-flavour production is due to four mechanisms, the first 
hard 2$\rightarrow$2 hard process, the subsequent hard process in MPI, 
gluon splitting from hard process, and from initial/final state gluon 
radiation (ISR/FSR). The dominating contribution in 
\mbox{PYTHIA 8} at LHC energies is the ISR/FSR mechanism which 
contributes 62\% and 40\% for D and B-meson production, respectively.
The EPOS model incorporates initial conditions according to the 
Gribov-Regge multiple scattering framework. Individual 
scatterings are initiated by Pomerons, and are associated to parton 
ladders. Each of these parton ladders represents a hard pQCD process 
with initial and final state radiation. A saturation scale is introduced 
for taking account of non-linear effects. With these initial conditions, 
a hydrodynamical evolution can be defined on the core of the collision. 
The EPOS calculation without hydro-evolution predicts an approximate linear 
increase of D-meson production  as function of relative charged-particle 
multiplicity. The EPOS model with hydro-evolution results in a deviation 
from linear dependence. Within  the percolation model, high-energy hadronic 
collisions are dominated by the exchange of colour sources of finite spatial 
dimension between projectile and target \cite{percol}. The percolation model 
predicts a faster than linear increase of the relative D-meson yield.

\section{Proton-Lead Collisions}
\label{sec-6}

The analysis of the p-Pb collision system sheds light on the role of initial 
and final-state effects when studying differences between the pp and Pb-Pb 
systems \cite{pp_PbPb}. The nuclear shadowing of PDFs, the existence of CGC in 
the initial state and the hydrodynamic phase in the evolution of the system, 
for example, affect bulk observables quite differently in the pp, p-Pb 
and Pb-Pb collision systems.
  
\begin{figure}[h!]
\begin{subfigure}[t]{0.48\textwidth}
\includegraphics[width=6.2cm]{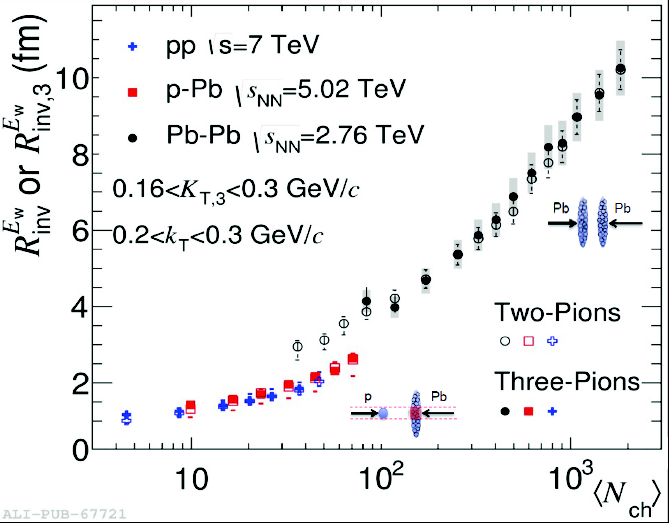}
\put(-178.,-2.2){\colorbox{white}{\makebox(174.,.1){ }}}
\put(-1.2,0.){\colorbox{white}{\makebox(.1,136.){ }}}
\end{subfigure}
\begin{subfigure}[ht]{0.50\textwidth}
\vspace{-4.6cm}
\includegraphics[width=6.cm]{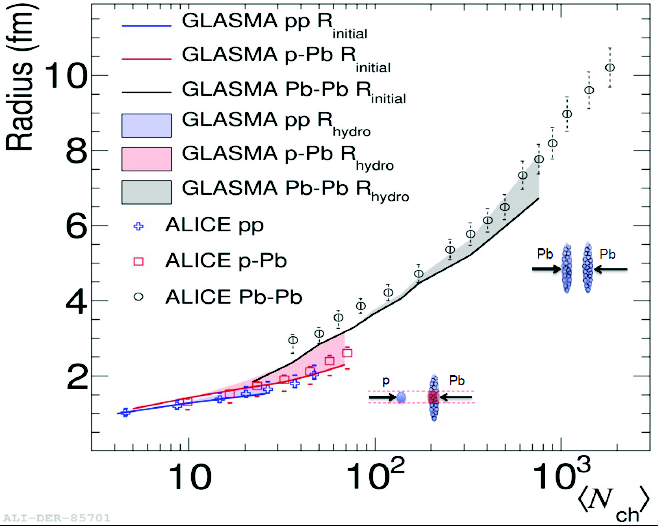}
\put(-153.2,16.2){\bf 0}
\put(-172.,-2.3){\colorbox{white}{\makebox(168.,.1){ }}}
\end{subfigure}
\caption{Pion source sizes are shown on the left, 
model calculations on the right.}
\label{fig-6}       
\end{figure}

The source sizes derived from pion correlations are shown on the left in
Fig. \ref{fig-6} as a function of the charged particle multiplicity for the 
pp, p-Pb and Pb-Pb systems \cite{source_pp_PbPb}. Here, $R_{inv}$ expresses the 
invariant source radius derived by a Gaussian fit to the two-pion correlation 
function. The quantity $R_{inv,3}$  denotes the corresponding quantity as 
determined by analysis of three-pion cumulant distributions. At the same 
multiplicities within the range $60 < N_{ch}< 80$, the source radii in the pp
and p-Pb systems show similar values, with a significantly larger value in the 
Pb-Pb system reflecting the hydrodynamic expansion stage of a heavy-ion 
collision.  These ALICE data are compared on the right of Fig. \ref{fig-6} to  
predictions by the CGC based IP-Glasma model \cite{source_CGC}.

\section{Lead-Lead Collisions}
\label{sec-7}

\subsection{Exploring the QCD Phase Diagram with Heavy-Ion Collisions}

Collisions of heavy-ions at high energies produce QCD matter at extreme 
values of energy densities and temperatures \cite{whitepaper_HI}. Lattice 
gauge calculations indicate that a QCD phase transition takes place under 
such conditions. In lattice QCD approach, the partition function of the 
grand canonical ensemble is evaluated stochastically by Monte Carlo sampling 
of field configurations. Thermodynamic state functions such as pressure and 
energy densities can subsequently be evaluated. The transition from 
hadronic matter to a state of deconfined quarks and gluons, the 
Quark-Gluon Plasma (QGP) is seen in such calculations at a critical 
temperature of T$_{\text{cr}} \sim$ 160 MeV. At this critical temperature, 
the chiral condensate $<\!\!\bar{q}q\!\!>$ is rapidly diminishing, thereby 
indicating the restoration of broken chiral symmetry \cite{chiralsym}.  
A multitude of experimental observables are analyzed to characterize the 
nature of the QCD phase diagram, and to establish signals of partial 
chiral symmetry restoration. The present understanding of this deconfined 
state is a strongly coupled QGP which behaves as a liquid with a very low 
viscosity to entropy ratio \cite{QGP}.  

\subsection{Soft Probes in Heavy-Ion Collisions}

The analysis of the azimuthal anisotropy of particle production in 
heavy-ion collisions reveals information on the initial geometry  of the 
overlap zone of the colliding nuclei, on the equation of state of the
produced QCD medium and its transport properties. 

\vspace{-0.2cm}
\begin{figure}[h!]
\begin{subfigure}[t]{0.48\textwidth}
\includegraphics[width=5.4cm]{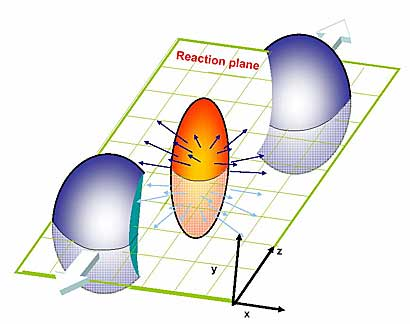}
\end{subfigure}
\begin{subfigure}[ht]{0.50\textwidth}
\vspace{-3.4cm}
\includegraphics[width=6.cm]{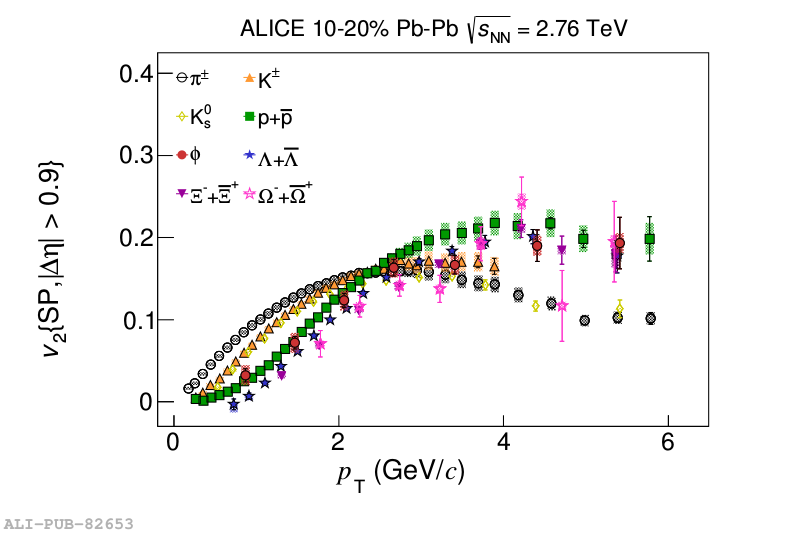}
\end{subfigure}
\caption{Heavy-ion collision geometry is shown on the left, $\nu_{2}$-flow
coefficients on the right.}
\label{fig-7}       
\end{figure}

\vspace{-0.4cm}
On the left side of Fig. \ref{fig-7}, a heavy-ion collision geometry is
shown. The azimuthal distribution of particles produced in the almond-shape 
overlap zone is analyzed with respect to the reaction plane by

\begin{equation}
\footnotesize \frac{d^{2}N}{p_{T}dp_{T}d\phi} \propto 1 + 
2\sum\limits_{n=1}^{\infty}\nu_{n}(p_{T})\; \text{cos}[n(\phi-\Psi_{n})] 
\label{eq-1}
\end{equation}

The azimuthal particle distribution of Eq. \ref{eq-1} is characterized by 
the symmetry planes $\Psi_{n}$ and by the coefficients $\nu_{n}$.
Such distributions test hydrodynamical models, and allow to extract 
properties of the QCD medium such as viscosity.
   
On the right side of Fig. \ref{fig-7}, the $\nu_{2}$-coefficient of the
Fourier decomposition of the azimuthal anisotropy is shown for the centrality 
class 10\%-20\% \cite{Alice_v2}. A distinctive mass ordering of the flow
coefficient $\nu_{2}$ is seen at low values of transverse momentum p$_{T}$.
This mass ordering is attributed to the interplay between elliptic and 
radial flow. Radial flow tends to deplete the spectrum at 
\mbox{low values of p$_{T}$,} which increases with increasing particle mass 
and transverse velocity. In a system with azimuthal anisotropy, this 
depletion is larger in-plane than out-of-plane, thereby reducing $\nu_{2}$. 
At large values of p$_{T}$, the flow coefficients $\nu_{2}$ show a tendency 
to order according to the particle type of baryon and meson.
Such scaling was first observed at RHIC, and was interpreted as evidence
that quark degrees of freedom dominate in the early stages of heavy-ion
collisions when collective flow develops \cite{RHIC}.

\subsection{Hard Probes in Heavy-Ion Collisions}

\begin{figure}[h!]
\begin{subfigure}[t]{0.48\textwidth}
\includegraphics[width=5.6cm]{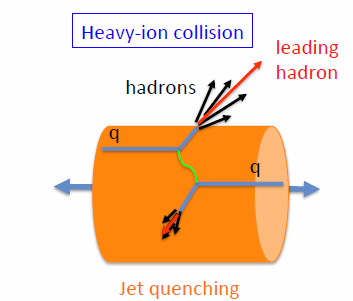}
\end{subfigure}
\begin{subfigure}[ht]{0.50\textwidth}
\vspace{-3.0cm}
\includegraphics[width=5.6cm]{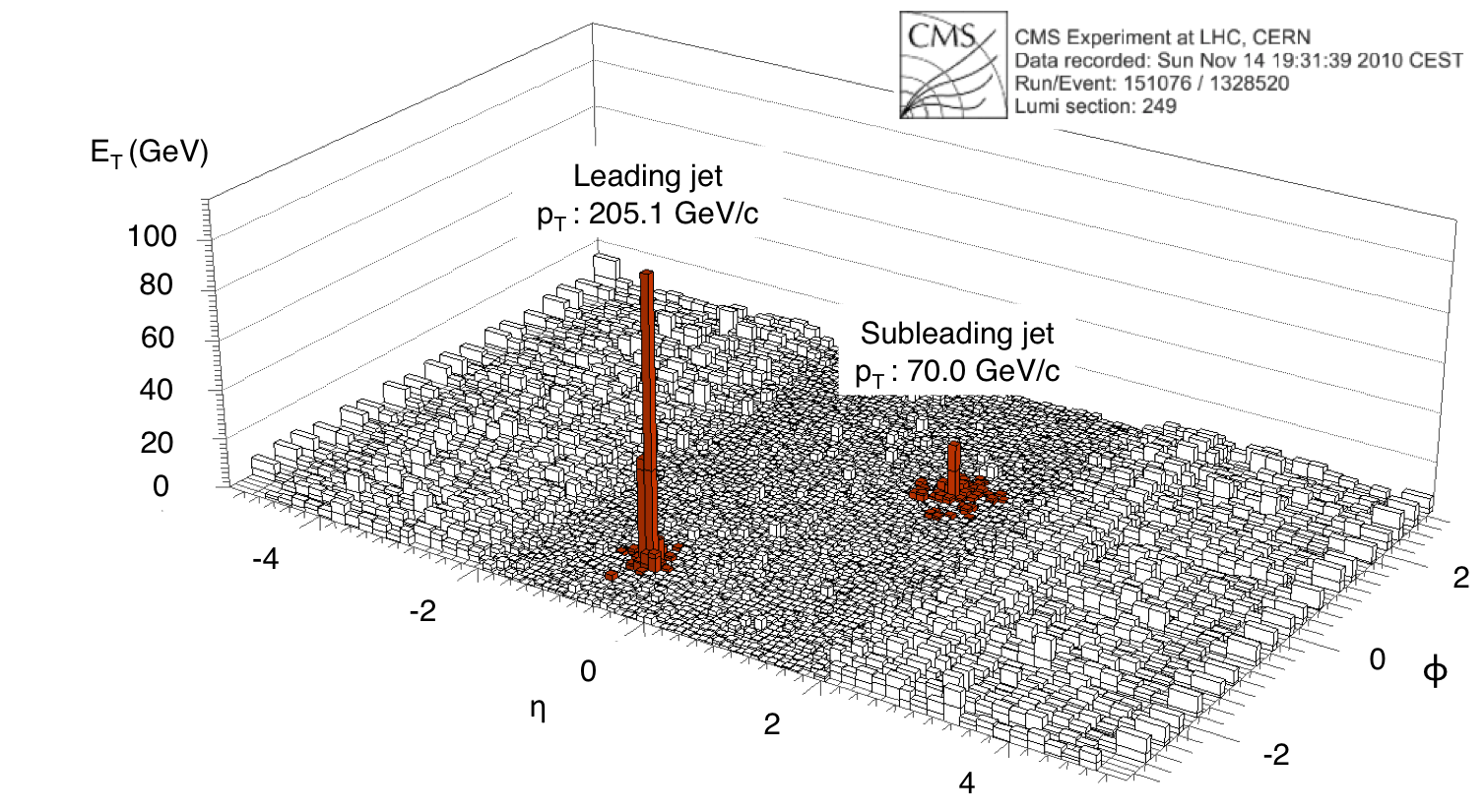}
\end{subfigure}
\caption{Hard parton scattering is shown on the left, a dijet event
on the right (figure from Ref. \cite{CMS_jet}).}
\label{fig-8}       
\end{figure}

The analysis of jets produced in scatterings of highly virtual 
quarks or gluons provides an important tool to test the  Standard Model
in the perturbative region of QCD. In heavy-ion collisions,  the analysis 
of jet quenchinq  allows to study the QCD medium produced in the collision.
In Fig. \ref{fig-8} on the left, a  hard parton scattering is shown.  
The hadronization of the two partons after scattering results
in two jets emerging on opposite sides in azimuthal angle. 
Of particular interest are hadron-photon and jet-photon correlations.
The photon is expected to escape the interaction zone with much
reduced interaction as compared to hadrons, and hence carries
the information of the hard-scattering kinematics. 

On the right of Fig. \ref{fig-8}, a dijet event is shown with the leading jet
of transverse momentum \mbox{p$_{T}$ = 205.1 GeV/c,} and the subleading jet
of  p$_{T}$ = 70.0 GeV/c at the opposite azimuthal angle \cite{CMS_jet}.  

\vspace{-0.2cm}

\vspace{-0.4cm}
The energy of the partons of the jet is reduced in heavy-ion collisions as 
compared to pp collisions due to medium-induced gluon radiation and collisional
energy loss. In perturbative QCD, the production cross section of the initial 
hard scattered parton can be calculated. The hadronization of the initial 
scattered parton can be calibrated by jet measurements in pp collisions. 
Jet quenching in heavy-ion collisions can be analyzed by studying whether 
these jet spectra can be understood as an incoherent superposition of 
nucleon-nucleon collisions. The analysis of ALICE data reveals that jets with 
transverse momenta $40 < p_{T,jet} < 120$ GeV/c in heavy-ion collisions are 
strongly suppressed in the 10\% most central events \cite{Alice_jet}.

\subsection{Low Mass Dileptons}

In the limit of vanishing quark masses, the QCD Lagrangian contains
a symmetry due to the conserved right or left-handedness of the 
zero-mass spin-1/2 particles. This chiral symmetry is spontaneously
broken, leading to a population of the  QCD ground state by scalar 
quark-antiquark pairs \cite{chiralsym_break}.
In the QCD medium formed in  heavy-ion collisions at high energies,
the expected partial restoration of chiral symmetry is reflected by 
significant modifications of hadron properties. 
Of particular importance for the study of such in-medium modifications
is the $\rho$-meson. Due to the short life-time \mbox{of $c\tau$= 1.3 fm,}
$\rho$-decays occur predominantly within the QCD-medium formed
in the heavy-ion collision. The measurement of the in-medium $\rho$-spectral 
function through the $\rho$-dilepton decays allows to test the nature of 
the formed QCD medium. Both  an in-medium mass shift, as well as a broadening 
of the $\rho$-meson have been advocated \cite{BrownRho,Rapp}.

\begin{figure}[h!]
\hspace{1.cm}
\begin{subfigure}[t]{0.48\textwidth}
\includegraphics[width=5.6cm]{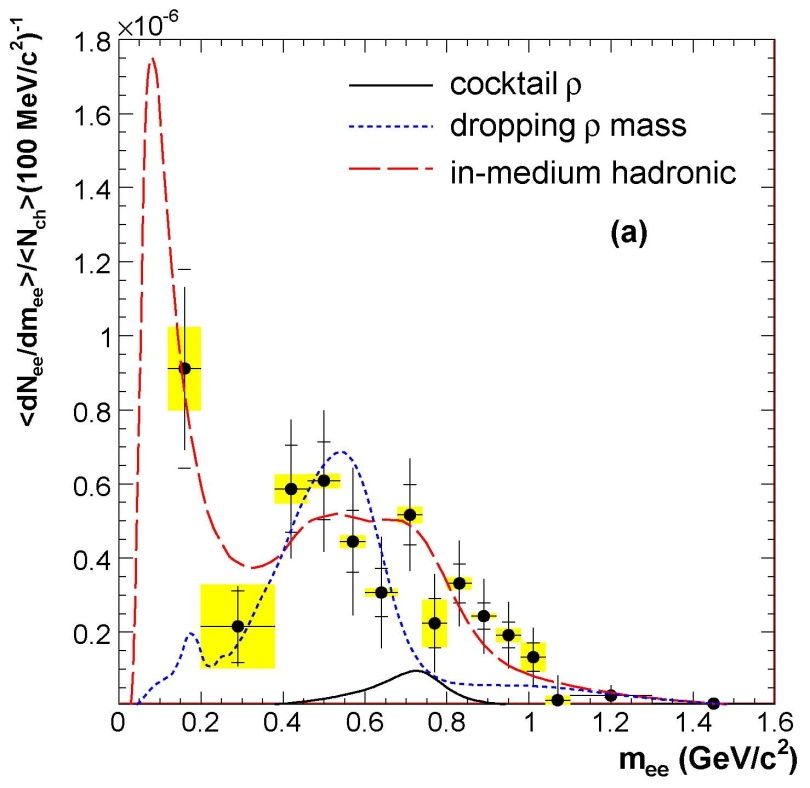}
\end{subfigure}
\begin{subfigure}[ht]{0.50\textwidth}
\vspace{-5.2cm}
\includegraphics[width=4.4cm]{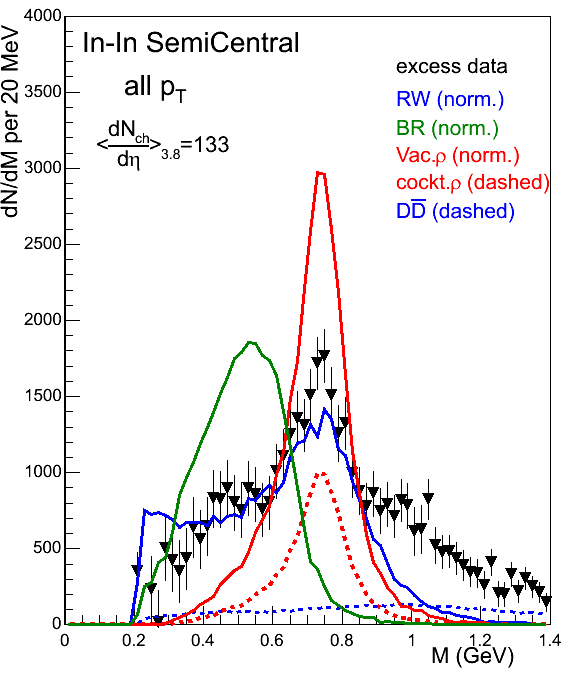}
\end{subfigure}
\caption{Invariant dilepton mass spectra measured by CERES (left)
and NA60 (right).}
\label{fig-9}       
\end{figure}

In Fig. \ref{fig-9} on the left, the invariant mass distribution
of $e^{+}e^{-}$-pairs measured by the CERES collaboration in Pb-Au
collisions at 158A GeV/c is shown \cite{Ceres_rho}.
The data points represent the $e^{+}e^{-}$-pair yield after subtraction 
of the hadronic cocktail, i.e. the contribution to this yield from 
known sources such as $\pi^{0},\eta,\eta'$ and $\omega$ Dalitz-decays.
In this figure, the dotted and long-dashed lines represent the 
expectations as derived from the dropping and broadening $\rho$-mass 
scenario, respectively. Within the statistics of these data, the dropping 
$\rho$-mass scenario is disfavored. On the right of Fig. \ref{fig-9}, the 
excess of the $\mu^{+}\mu^{-}$-pair yield measured by the NA60 collaboration 
in In-In collisions \mbox{is shown \cite{NA60_rho}.}
These data are compared to expectations from the dropping $\rho$-mass
scenario (green line) and  the broadening $\rho$-mass scenario (blue line).
The solid red line represents the expectation of this yield in case of 
no changes to the $\rho$-spectral function.
The dashed red line is shown for illustration, and indicates the expected
contribution from the cocktail $\rho$ bound  by the ratio of 
$\rho/\omega$ = 1.2 measured  at high \mbox{p$_{T} >$ 1.6 GeV/c.}
From these data, an unmodified $\rho$ is clearly ruled out, as well 
as the dropping mass scenario. The most realistic description of 
these data is within the  broadening $\rho$-mass scenario.

The ALICE collaboration has measured the yield of low-mass 
dileptons at  pseudorapidities \mbox{$|\eta| < 0.8 $} in Run I of the LHC
in pp collisions at $\sqrt{s}$ = 7 TeV, 
and in p-Pb collisions at \mbox{$\sqrt{s_{\text{NN}}}$ = 5.02 TeV.}

\begin{figure}[h!]
\begin{subfigure}[t]{0.48\textwidth}
\includegraphics[width=8.4cm]{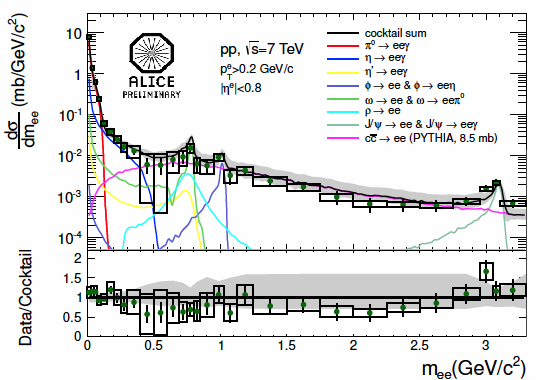}
\end{subfigure}
\hspace{1.4cm}
\begin{subfigure}[ht]{0.50\textwidth}
\vspace{-5.8cm}
\includegraphics[width=5.7cm]{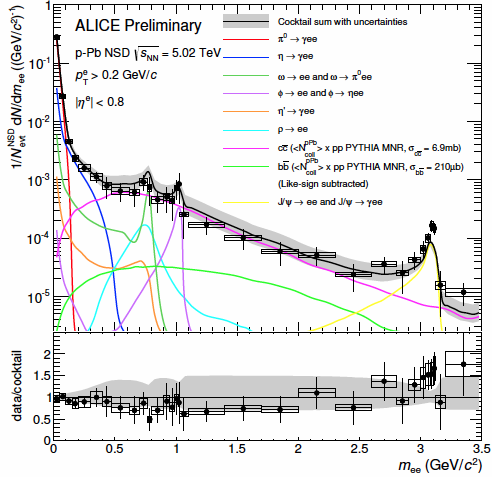}
\end{subfigure}
\caption{ALICE mass spectra of $e^{+}e^{-}$-pairs
in pp (left) and p-Pb collisions (right).}
\label{fig-10}       
\end{figure}

The differential cross section for $e^{+}e^{-}$-pair production in pp collisions
is shown in Fig. \ref{fig-10} on the left. Superimposed to the data
are the cocktail contributions expected from 
$\pi^{0},\eta,\eta',\omega,\phi$ and $J/\Psi$ Dalitz-decays, as well as 
from the two-body $e^{+}e^{-}$-decays of $\rho,\omega,\phi$ and $J/\Psi$.
At pair masses beyond the $\phi$-mass, the spectrum is dominated
by the contribution of $c\bar{c}$ decays.
The expected cocktail contribution is consistent with the data
up to a mass of m=3.5 GeV/c$^{2}$ as shown in the lower part on the left.
The differential cross section for $e^{+}e^{-}$-pair production in p-Pb 
collisions is shown in Fig. \ref{fig-10} on the right, together with the
expected cocktail contributions. The expected cocktail contribution is 
consistent with the data up to a mass of m=3.5 GeV/c$^{2}$ as shown in the 
lower part on the right.

The dilepton spectra measured in pp and p-Pb collisions in ALICE
serve as baseline to the corresponding measurements in Pb-Pb collisions.
Besides the in-medium modification of low-mass vector mesons discussed
above, the analysis of virtual direct photons in the low mass region 
and signatures of thermal radiation from the QGP in the intermediate 
mass region are of high interest. Such studies can be pursued in ALICE
by analysing dielectrons in the central barrel as well as by
examining dimuons in the ALICE muon spectrometer.

\section{The Upgrade Programme of ALICE}
\label{sec-8}

The ALICE detector upgrade programme is driven by the future ALICE physics 
plans \cite{Alice_up}. These plans center on rare physics probes where the 
ALICE capabilities on measuring low-p$_{T}$ tracks, in conjunction 
with excellent particle identification, are unique. 
Such probes are, for example, heavy flavour hadrons and quarkonia
at low p$_{T}$, low mass lepton pairs, jets, 
light nuclei and hypernuclei. These upgrade plans are formulated
under the assumption that the LHC will increase the Pb-beam intensity 
such that an interaction rate of 50 kHz is reached in Run III, corresponding 
to an instantaneous luminosity of L = 6 x 10$^{27}$cm$^{-2}$s$^{-1}$. Standard 
trigger strategies cannot be applied in most cases of the above-mentioned 
rare probes. The ALICE detectors hence need to be modified such 
that all interactions can be inspected. Based on these considerations, 
the ALICE detector upgrade programme for Run III starting in the year 2021
consists of

\begin{itemize}  

\item Upgrade of the TPC in order to operate in ungated mode 
to have a dead-time free readout

\item A new high-resolution, thin-material Inner Tracking System (ITS).
The resolution on distance of closest approach will be improved by a factor
of about 3 as compared to the present performance

\item Upgrade of the readout electronics of the TRD, TOF, PHOS detectors 
and the muon spectrometer for the future high-rate data taking  

\item Upgrade of the data acquisition and the high-level trigger system
for the future high-rate data taking

\item Upgrade of the offline processing software in order to handle the much
larger number of events

\end{itemize}

\section{Summary and Outlook}
\label{sec-9}

The ALICE collaboration has taken data in Run I and Run II of the LHC in
pp, p-Pb and Pb-Pb collisions. A multitude of physics analyses
of the  Pb-Pb data addresses the nature  of the QCD phase diagram.
The study of pp collisions addresses a variety of questions
relevant for an improved understanding of the non-perturbative sector of QCD.
The analyses of  p-Pb collisions reveal effects of cold-matter
initial and final-state effects in heavy-ion induced reactions.
The future of the ALICE physics programme is centered on measuring
rare probes with much improved statistics as is presently available.
The ALICE detectors are correspondingly upgraded in order to meet 
the increased data rate in the future.

\section*{Acknowledgements}

This work is supported by the German Federal Ministry of Education and 
Research under promotional reference 05P15VHCA1.

%
%

\end{document}